\documentclass[runningheads]{llncs}
\usepackage[T1]{fontenc}
\usepackage{graphicx}

\usepackage{amsmath}
\usepackage{lipsum}
\usepackage{graphicx}
\usepackage{subcaption}
\usepackage{multirow}
\usepackage{booktabs}
\usepackage{amsfonts}

\begin{document}
\title{Rethinking Annotator Simulation: \\ Realistic Evaluation of Whole-Body PET Lesion Interactive Segmentation Methods}
\titlerunning{Rethinking Annotator Simulation for Interactive PET Lesion Segmentation}
%
\author{Zdravko Marinov\inst{1,2} \and
Moon Kim\inst{3} \and
Jens Kleesiek\inst{3,4} \and
Rainer Stiefelhagen\inst{1}
}
%
\authorrunning{Z. Marinov et al.}
%
\institute{Karlsruhe Institute of Technology, Karlsruhe, Germany \\
\and
HIDSS4Health - Helmholtz Information and Data Science School for Health,
Karlsruhe/Heidelberg, Germany \\ 
\and 
Institute for AI in Medicine, University Hospital Essen, Essen, Germany \\ 
\and
Cancer Research Center Cologne Essen (CCCE), University Medicine Essen, Essen,\\
\email{$^1$\{firstname.lastname@kit.edu\} , $^3$\{firstname.lastname@uk-essen.de\}}
}
\maketitle              
\begin{abstract}


Interactive segmentation plays a crucial role in accelerating the annotation, particularly in domains requiring specialized expertise such as nuclear medicine. For example, annotating lesions in whole-body Positron Emission Tomography (PET) images can require over an hour per volume. While previous works evaluate interactive segmentation models through either real user studies or simulated annotators, both approaches present challenges. Real user studies are expensive and often limited in scale, while simulated annotators, also known as robot users, tend to overestimate model performance due to their idealized nature. To address these limitations, we introduce four evaluation metrics that quantify the user shift between real and simulated annotators. In an initial user study involving four annotators, we assess existing robot users using our proposed metrics and find that robot users significantly deviate in performance and annotation behavior compared to real annotators. Based on these findings, we propose a more realistic robot user that reduces the user shift by incorporating human factors such as click variation and inter-annotator disagreement. We validate our robot user in a second user study, involving four other annotators, and show it consistently reduces the simulated-to-real user shift compared to traditional robot users. By employing our robot user, we can conduct more large-scale and cost-efficient evaluations of interactive segmentation models, while preserving the fidelity of real user studies. Our implementation is based on MONAI Label and will be made publicly available.

\end{abstract}
%
%
\section{Introduction}
Deep learning models have made significant progress in segmenting anatomical structures and lesions in medical images but often rely on manually labeled datasets \cite{autopet,brats,msd,amos,total,isles}. This poses a challenge for volumetric medical data where annotating each voxel demands considerable time and expertise. Interactive segmentation mitigates this issue by leveraging less demanding annotations, such as clicks, instead of dense voxelwise labels \cite{review1,review2,swfastedit,petctgui,adaptive,deepigeos,mideepseg,BIFSeg,guiding}. Clicks are combined with the image as a joint input for the interactive model and guide it spatially toward the segmentation target. Annotators can refine model outputs by placing clicks in missegmented areas, leading to an improved segmentation and high-quality predictions \cite{swfastedit,petctgui,adaptive,deepigeos,mideepseg,BIFSeg,guiding}. Once approved by medical experts, these predictions may serve as new labels \cite{review1}. Prior methods evaluate interactive models by simulating clicks on the test split (a "robot user") \cite{guiding,moschidis,amrehn,kohli} or by involving real annotators in a user study \cite{swfastedit,petctgui,adaptive}. However, real user studies are costly, with a limited sample size, and robot users often overestimate model performance due to their idealized nature. Similar to a domain shift encountered when assessing models with out-of-domain data (e.g., from a different scanner), a \textit{user shift} arises when validating an interactive model via simulated robot users and deploying it in real clinical settings, where its performance often diverges \cite{review1}. We address these challenges for whole-body PET lesion segmentation with the following contributions:
\begin{enumerate}
    \item We evaluate 4 robot users \textbf{(R1)--(R4)} on the AutoPET dataset \cite{autopet} and conduct 2 user studies, each with 4 medical annotators, to show the disparity between simulated and real user performance of existing robot users.
    \item We introduce 4 evaluation metrics \textbf{(M1)--(M4)} to quantify the simulated-to-real user shift in terms of segmentation accuracy, annotator behavior, and conformity to ground-truth labels.  
    \item We propose a novel robot user that mitigates the pitfalls identified in 1. by simulating clicks that disagree with the ground-truth labels. Our robot user reduces the user shift (defined in 2.) and the segmentation performance gap to real users compared to previous robot users in both our user studies.
\end{enumerate}
\textbf{Related Work.}
Previous research on robot users mainly explores classical non-deep learning methods and overlooks evaluating the disparity with real annotators. For example, Kohli et al. \cite{kohli} compare four Graph Cut-based interactive models \cite{graphcut} and conclude that placing clicks at the center of the largest error consistently yields optimal results across all models. However, their comparison is limited to natural images, and they do not explore deep learning-based approaches. 
Moschidis and Graham \cite{moschidis} compare two robot users for 3D medical image segmentation: one targeting central regions and the other - boundary regions. However, their study also examines classical non-deep learning methods and lacks simulated clicks for iterative corrections.
Benenson et al. \cite{benenson} compare iterative boundary and central clicks, discovering that central clicks outperform boundary clicks, particularly when adding random noise perturbations, however, they also only explore the domain of natural images. The closest work to ours is Amrehn et al. \cite{amrehn}, which compares robot users using an interactive U-Net \cite{unet} for liver lesion segmentation. Their results suggest that a U-Net trained with a robot user using more spatially distributed clicks generalizes well when evaluated with a different robot user. However, they do not explore the generalization to real annotator interactions. In contrast to previous work, our focus lies on evaluating deep learning-based methods incorporating iterative corrections, with an emphasis on reducing the disparity between simulated and real annotators. Interactive segmentation reviews \cite{review1,review2} have discussed the lack of user-centric metrics for medical interactive segmentation. We address this by introducing 4 metrics that capture user behavior and quantify the simulated-to-real user shift.

\begin{figure}
    \centering
    \includegraphics[width=\linewidth]{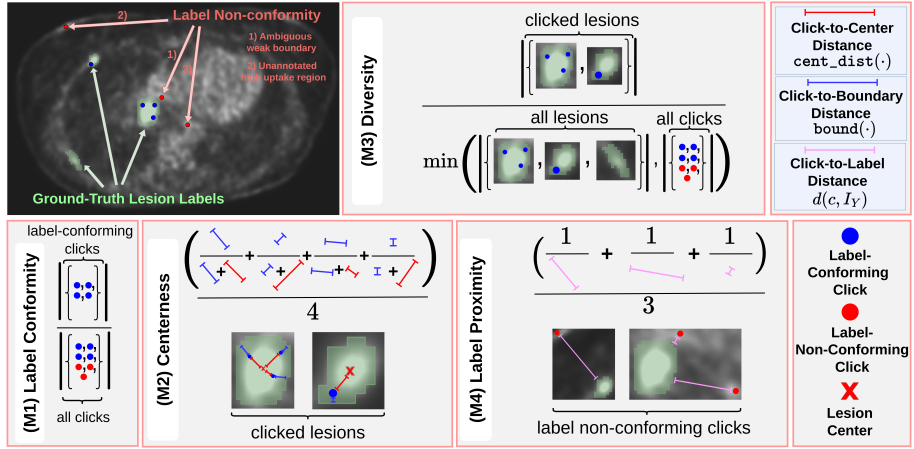}
    \caption{Our proposed evaluation metrics and examples of label non-conformity.}
    \label{fig:teaser}
\end{figure}
\vspace*{-1.1cm}

\section{Methods}
We explore iterative interactive models that simulate clicks in a loop of 10 iterations. In each click iteration $i \in \{1,...,10\}$, a robot user $R$ simulates a click, denoted as $\texttt{clicks}(R, I)[i] \in \mathbb{N}^3$, and combines it with the image $I \in \mathbb{R}^{W \times H \times D}$ as a joint input, where $W \times H \times D$ are the image dimensions. Using this joint input, the model predicts a segmentation mask $\texttt{pred}(I)[i] \in \{0,1\}^{W \times H \times D}$. Then, the missegmented regions within this prediction, denoted as $\texttt{err}(I)[i] \in \{0,1\}^{W \times H \times D}$, are employed to generate $\texttt{clicks}(R, I)[i+1]$ for the next iteration. We provide a notation table for all our equation terms in the supplementary.
\subsection{Robot Users}

\textbf{(R1) Center Click:} A common approach is to simulate clicks in the center of the largest missegmented component \cite{review1,center}. However, the first click is placed in the center of the largest component of the label $I_Y$. This is defined as:

\begin{equation}
\resizebox{0.85\linewidth}{!}{
    \label{eq:r1}
    $
    \texttt{clicks}(R1, I)[i] = 
    \begin{cases}
        \texttt{center}(\texttt{largest\_component}(I_Y)), & \text{if } i = 1 \\
        \texttt{center}(\texttt{largest\_component}(\texttt{err}(I)[i-1])), & \text{if } i > 1
    \end{cases}
    $
}
\end{equation}
where $I_Y \in \{0,1\}^{W \times H \times D}$ is the ground-truth label for image $I$, $\texttt{center}(\cdot)$ computes the geometric center of a component as in \cite{center}, and $\texttt{largest\_component}(\cdot)$ computes the largest connected component. 

\textbf{(R2) Uncertainty:} Zheng et al. \cite{uncertainty} sample a click in each iteration using the epistemic uncertainty of the model as a sampling distribution, defined as:

\begin{equation}
\resizebox{0.68\linewidth}{!}{
    \label{eq:r2}
     $\texttt{clicks}(R2, I)[i] \sim 
    \begin{cases}
      \texttt{uniform}(I_Y), & \text{if } i = 1 \\
        \texttt{epistemic}(\texttt{pred}(I)[i-1]) , & \text{if } i > 1 \\
    \end{cases}$
}
\end{equation}
where $\texttt{epistemic}(\cdot)$ is the normalized epistemic uncertainty in $[0,1]$ using Monte Carlo Dropout \cite{mcdropout}, and $\texttt{uniform}(X)$ defines a uniform distribution over the non-zero entries of $X$.

\textbf{(R3) Euclidean Distance Transform (EDT):} Previous methods \cite{swfastedit,petctgui} apply the EDT on missegmented regions as a sampling distribution for clicks:
\begin{equation}
\resizebox{0.525\linewidth}{!}{
    \label{eq:r3}
    $
    \texttt{clicks}(R3, I)[i] \sim 
    \begin{cases}
        \texttt{uniform}(I_Y), & \text{if } i = 1 \\
        \texttt{EDT}(\texttt{err}(I)[i-1]) , & \text{if } i > 1 \\
    \end{cases}$
}
\end{equation}
where \texttt{EDT}$(\texttt{err}(I)[i-1])$ is the normalized EDT of the non-zero entries in the missegmented regions $\texttt{err}(I)[i-1]$ from the previous iteration.

\textbf{(R4) Uniform:} The final robot user samples uniformly either from the previous error \cite{amrehn} or from the label for the first click as:
\begin{equation}
\resizebox{0.6\linewidth}{!}{
    \label{eq:r4}
    $
    \texttt{clicks}(R4, I)[i] \sim 
    \begin{cases}
        \texttt{uniform}(I_Y), & \text{if } i = 1 \\
        \texttt{uniform}(\texttt{err}(I)[i-1]) , & \text{if } i > 1 \\
    \end{cases}$
}
\end{equation}
\textit{\textbf{Note:}} In each iteration we simulate two types of clicks: $\texttt{clicks}(R, I)[i]^{\text{lesion}}$ and $\texttt{clicks}(R, I)[i]^{\text{background}}$. We designate the under- and over-segmented regions as missegmented areas $\texttt{err}(I)[i]$ for the "lesion" and "background" classes respectively, and omit the class labels in Eq.(\ref{eq:r1})-(\ref{eq:r4}), for clarity.

\textbf{$\textbf{(R}_{\textbf{ours}}$): Our Robot User:} 
In our first user study, we found that 25\% of our annotators' clicks are outside the ground-truth labels. Label non-conforming clicks stem from two factors (see Fig. \ref{fig:teaser}, top left): 1) ambiguous weak boundaries in the low-resolution PET scans, leading to clicks slightly outside the label boundaries; 2) and unannotated high uptake regions, spatially isolated from ground-truth labels. To address the first issue, we propose integrating click perturbations to spatially displace clicks with a probability $p_\text{perturb}$. For the second issue, we propose to systematically incorporate label non-conformity by sampling clicks in high uptake regions outside the ground-truth labels with a probability $p_\text{system}$. To achieve this, our robot user extends \textbf{(R1)} and is defined as:
\begin{equation}
\resizebox{0.8\linewidth}{!}{
$\texttt{clicks}(R_\text{ours}, I)[i] = 
\begin{cases}
\texttt{clicks}(R1, I)[i] & \text{if } p_{i,1} \geq p_\text{perturb} \text{ and } p_{i,2} \geq p_\text{system} \\
\texttt{clicks}(R1, I)[i] + \widetilde{z}, & \text{if } p_{i,1} < p_\text{perturb} \text{ and } p_{i,2} \geq p_\text{system}\\
\widetilde{s}, & \text{if } p_{i,1} \geq p_\text{perturb} \text{ and } p_{i,2} < p_\text{system} \\
\widetilde{s}  + \widetilde{z} , & \text{if } p_{i,1} < p_\text{perturb} \text{ and } p_{i,2} < p_\text{system} \\
\end{cases}$
}
\end{equation}
where $\widetilde{s} \sim \texttt{SUV}(I, I_Y)$ and $\widetilde{z} \sim \mathcal{U}_{[-a, a]^3}$. $\texttt{SUV}(I, I_Y)$ defines a distribution over the normalized Standardized Uptake Values in $I$ which are outside the label $I_Y$, $\widetilde{z}$ is a random perturbation with a maximal amplitude $a \in \mathbb{N}$, and each iteration $p_{i,1}$, $p_{i,2}$ are independently sampled from $\mathcal{U}_{[0,1]}$ to decide which case is applied.

\subsection{Model Architecture and Dataset}
We use the pre-trained SW-FastEdit \cite{swfastedit} interactive model based on MONAI Label \cite{monailabel} with a U-Net backbone \cite{unet} and conduct our user studies on the openly available AutoPET \cite{autopet} dataset which consists of 1014 PET/CT volumes with annotated tumor lesions of melanoma, lung cancer, or lymphoma. We exclusively utilize PET data and use SW-FastEdit's official test split of 10\% of the volumes. The PET volumes have a voxel size of $2.0 \times 2.0 \times 3.0\text{mm}^3$ and an average resolution of $400 \times 400 \times 352$ voxels. Both user studies were conducted using 3D Slicer \cite{slicer} and its MONAI Label plugin. We implemented our robot user experiments with MONAI Label \cite{monailabel} and will release the code.
\section{Experiments and Results}
\subsection{Evaluation Metrics}
For all metrics, we denote $\mathcal{I}$ as the set of PET images labeled in a user study, $\mathcal{A}$ as the set of real annotators participating in the study, and fix the number of clicks per image to 10. We visualize examples for \textbf{(M1)-(M4)} in Fig. \ref{fig:teaser}.

\textbf{(M1) The Label Conformity} for an annotator $A$ is defined as:

\begin{equation}
\resizebox{0.7\linewidth}{!}{
    $\textbf{M}_{1}(A) = \frac{1}{|\mathcal{I}|}\frac{1}{10}\sum_{I \in \mathcal{I}}\sum_{i=1}^{10} \textbf{[} I_Y[\small\texttt{clicks}(A, I)[i]]=1 \textbf{]}$ 
}
\end{equation}
where $\textbf{[}\cdot\textbf{]}$ is the Iverson bracket. \textbf{(M1)} measures to what extent an annotator's clicks agree with the ground-truth labels of the PET images.

\textbf{(M2) The Centerness} for annotator $A$ is defined as:
\begin{equation}
\resizebox{0.875\linewidth}{!}{
    $\textbf{M}_{2}(A)= \frac{1}{|\mathcal{I}|}\frac{1}{|\Bar{C}(A, I)|}\sum_{I \in \mathcal{I}}\sum_{c \in \Bar{C}(A, I)} \frac{\texttt{bound}(c, I_Y)}{\texttt{bound}(c, I_Y) + \texttt{cent\_dist}(c, I_Y)}$
}
\end{equation}
where $\Bar{C}(A, I) =\{c \ | \ c \in \texttt{clicks}(A, I) \ \text{and} \  I_Y[c]=1\}$ is the set of label conforming clicks of annotator $A$ for image $I$, $\texttt{bound}(c, I_Y)$ is the minimum distance of click $c$ to the boundary of the label $I_Y$, and $\texttt{cent\_dist}(c, I_Y)$ is the minimum distance of click $c$ to the center of the label $I_Y$. Small \textbf{(M2)} values indicate that label-conforming clicks are placed near the boundary, whereas large values show that clicks are placed near the central regions of the label.

\textbf{(M3)} \textbf{The Click Diversity} for annotator $A$ is defined as:
\begin{equation}
\resizebox{0.875\linewidth}{!}{
    $\textbf{M}_{3}(A) = \frac{1}{|\mathcal{I}|}\sum_{I \in \mathcal{I}}\frac{|\{\widetilde{Y} \ | \ \widetilde{Y} \in \texttt{components}(I_Y) \ \text{and} \  \exists c \in \texttt{clicks}(A, I) : \ c \in \widetilde{Y} \} |}{\min(|\texttt{components}(I_Y)|, \ |\texttt{clicks}(A, I)|)}$
}
\end{equation}
where $\texttt{components}(\cdot)$ is the set of all connected components. \textbf{(M3)} measures to what extent clicks are spread out in different connected components in the label. 

\textbf{(M4) The Label Proximity} for an annotator $A$ is defined as:

\begin{equation}
\resizebox{0.625\linewidth}{!}{
    $\textbf{M}_{4}(A) = \frac{1}{|\mathcal{I}|}\frac{1}{|\hat{C}(A,I)|}\sum_{I \in \mathcal{I}}\sum_{c\in\hat{C}(A, I)} \frac{1}{d(c, I_Y)}$
}
\end{equation}
where $\hat{C}(A, I)=\{c \ | \ c \in \texttt{clicks}(A, I) \ \text{and} \ I_Y[c] = 0\}$ is the set of label non-conforming clicks of annotator $A$ for image $I$, and $d(c, I_Y)=\min(\{||c - y|| \ | \ y \in \mathbb{N}^{W \times H \times D} \ \text{and} \ I_Y[y]=1\})$. \textbf{(M4)} computes the average inverse distance of the annotator clicks outside the ground-truth label to the label $I_Y$. Higher \textbf{(M4)} values suggest non-conforming clicks are close to the label boundary, while lower values indicate clicks are far from any component of the label $I_Y$, suggesting systematic non-conformity.

\textbf{(M5) The Consistent Improvement} is defined in \cite{guiding} as:
\begin{equation}
\resizebox{0.8\linewidth}{!}{
    $\textbf{M}_{5}(A) = \frac{1}{|\mathcal{I}|}\frac{1}{10} \sum_{I \in \mathcal{I}}\sum_{i=1}^{10} \textbf{[}\texttt{dice}(A, I)[i] > \texttt{dice}(A, I)[i-1]\textbf{]}$
}
\end{equation}
where $\texttt{dice}(A, I)[i]$ is the Dice score after annotator $A$'s $i^{\text{th}}$ click on image $I$.

\textbf{(M6) The User Shift} determines the mean absolute difference in all metrics \textbf{(M1)-(M5)} between a simulated robot user $R$ and all real annotators $\mathcal{A}$:
\begin{equation}
\resizebox{0.65\linewidth}{!}{
    $\textbf{M}_{6}(R, \mathcal{A}) = \frac{1}{|\mathcal{A}|}\frac{1}{5}\sum_{A \in \mathcal{A}} \sum_{i=1}^5|\textbf{M}_i(R) - \textbf{M}_i(A)|$
}
\end{equation}

\textbf{(M7) The Dice Difference} for a robot user $R$ is defined as: 
\begin{equation}
\resizebox{0.85\linewidth}{!}{
$\textbf{M}_{7}(R, \mathcal{A}) = \frac{1}{|\mathcal{I}|}\frac{1}{|\mathcal{A}|}\frac{1}{10}\sum_{I \in \mathcal{I}} \sum_{A \in \mathcal{A}}\sum_{i=1}^{10} |\texttt{dice}(A, I)[i] - \texttt{dice}(R, I)[i]|$ 
}
\end{equation}

\textbf{(M6)} quantifies the fidelity of the robot user in emulating annotator behavior, while \textbf{(M7)} evaluates its ability to reproduce the segmentation performance of the interactive model as used by real annotators.
\subsection{User Studies and Results}
\textbf{Setup.} We conduct two user studies, each with four annotators from a medical background. In both studies, annotators were instructed to place 10 "lesion" and 10 "background" clicks, updating the model prediction after each pair of clicks to replicate the workflow of simulated robot users. In our first user study, four annotators labeled the same 10 PET volumes from the test split. We used this user study to determine the optimal values of $p_\text{perturb}$ and $p_\text{system}$ for our robot user. In our second user study, four different annotators labeled 6 PET volumes. We conducted this as a "validation" user study to confirm that our results from the first user study generalize to other volumes and annotators.
For both studies, we applied each robot user to the same PET images annotated by the real users.

\begin{figure}[h]
\hspace*{-0.1cm}
  \begin{subfigure}[b]{0.375\textwidth}
    \includegraphics[width=1.4\textwidth]{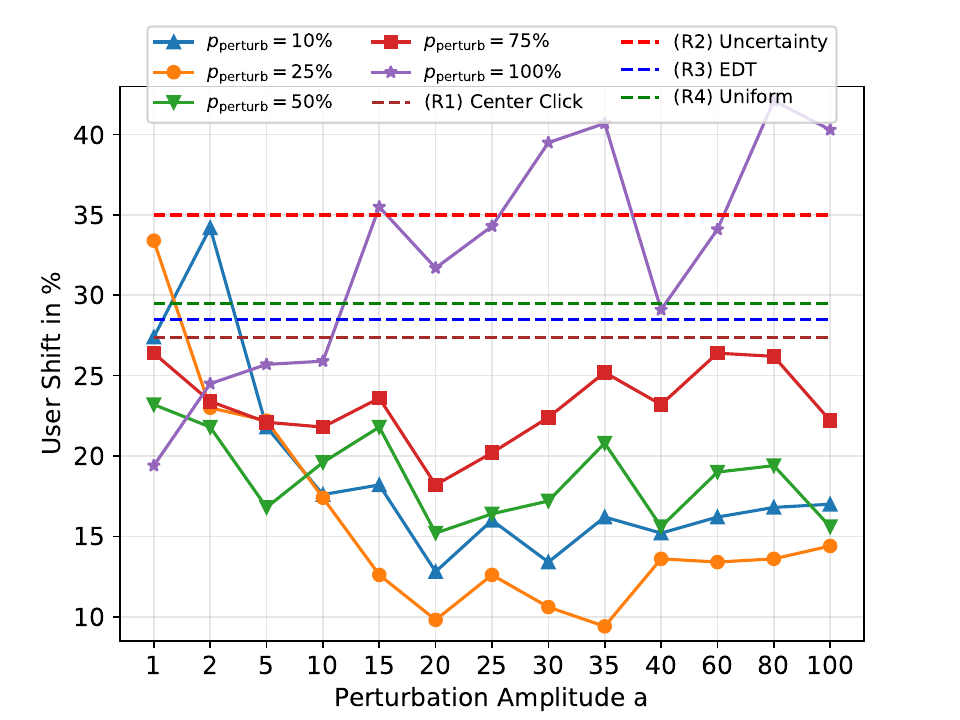}
  \end{subfigure}
  \hspace*{1.1cm}
  \begin{subfigure}[b]{0.375\textwidth}
    \includegraphics[width=1.4\textwidth]{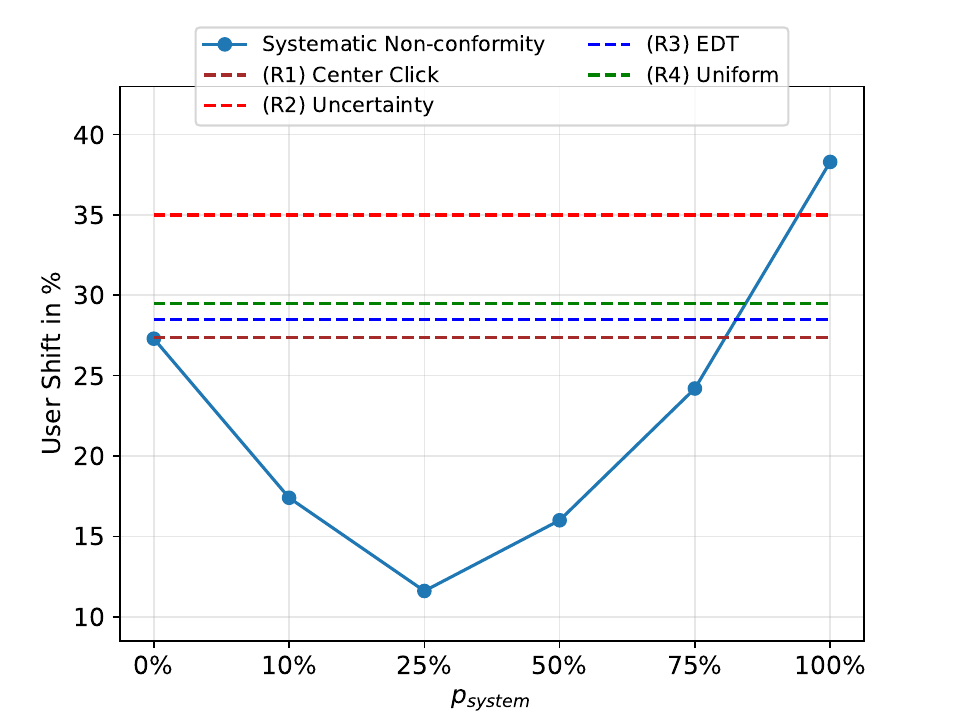}
  \end{subfigure}
  \caption{Analysis of $p_\text{perturb}$ (left) and $p_\text{system}$ (right) in our first user study.}
  \label{fig:ablation}
\end{figure}

\begin{table}[h]
    \centering
    \caption{User Shift and Dice Difference of all robot users on both user studies.}
\renewcommand{\arraystretch}{1.2} 
\scalebox{0.65}{
    \begin{tabular}{ll|cccc|cccccc}
    \toprule
        {} & {} & \multicolumn{4}{c}{Previous Work}  & \multicolumn{6}{c}{Ours ($a=35$)} \\ \cmidrule(lr){3-12}
        {} & {} & \multirow{2}{0.8cm}{\centering \textbf{(R1)}} & \multirow{2}{0.8cm}{\centering \textbf{(R2)}} & \multirow{2}{0.8cm}{\centering \textbf{(R3)}} & \multirow{2}{0.8cm}{\centering \textbf{(R4)}} & $p_\text{perturb}$ & $25\%$ & $19.6\%$ & 
        $13.4\%$ &
        $6.7\%$ &
        $0\%$ \\ 
        {} & {} & {} & {} & {} & {} & $p_\text{system}$ & $0\%$ & $6.7\%$ & 
        $13.4\%$ &
        $19.6\%$ &
        $25\%$ \\ 
        \hline
        \multirow{2}{*}{\rotatebox[origin=c]{0}{\shortstack{\text{User} \text{Study 1}}}} & \textbf{(M6)} User Shift & $27.4$ & $35.0$ & $28.5$ & $29.5$ & {} & $9.4$ & $8.4$ & \textbf{6.8} & $9.0$ & $11.6$  \\
        {} & \textbf{(M7)} Dice Difference & $8.7$ & $10.0$ & $9.2$ & $11.6$ & {} & $6.0$ & $5.3$ & \textbf{3.6} & $5.8$ & $6.9$\\  
        \hline
        \multirow{2}{*}{\rotatebox[origin=c]{0}{\shortstack{\text{User} \text{Study 2}}}}  & \textbf{(M6)} User Shift & $30.0$ & $31.7$ & $33.8$ & $30.0$ & {} & 8.4 & 7.6 & \textbf{6.7} & 8.6 & 9.2 \\
        {} & \textbf{(M7)} Dice Difference & $8.5$ & $9.0$ & $7.0$ & $7.5$ & {} & 5.3 & 4.8 & \textbf{3.7} & 6.2 & 6.7 \\
    \bottomrule
    \end{tabular}}
    \label{tab:results}
\end{table}
\vspace*{-0.75cm}

\textbf{Results: Our Robot User.} In the first user study, we assessed our robot user by varying $p_\text{perturb}$, $p_\text{system}$ and the perturbation amplitude $a$ and plotted the results in Fig. \ref{fig:ablation}. 
Spatial perturbations with $p_\text{perturb} \leq 75\%$ consistently outperform existing robot users in terms of user shift. The optimal user shift is achieved with $p_\text{perturb} \leq 75\%$ and $a \in [20,35]$, in particular with $p_\text{perturb} = 25\%$ and $a = 35$, deteriorating with $a > 35$ or $p_\text{perturb} = 100\%$ due to the excessive spatial noise. Incorporating systematic non-conformity also consistently reduces the user shift, with $p_\text{system}=25\%$ as the optimal value, similar to $p_\text{perturb}$. Since $25\%$ is the optimal value for both $p_\text{perturb}$ and $p_\text{system}$, we explore mixing them with a joint probability of $25\%$. The results in Table \ref{tab:results} show that mixing further reduces the user shift as well as the Dice difference, leading to optimal results when $p_\text{system}=p_\text{perturb}$.

\begin{figure}[b]
    \centering
    \includegraphics[width=0.7\textwidth]{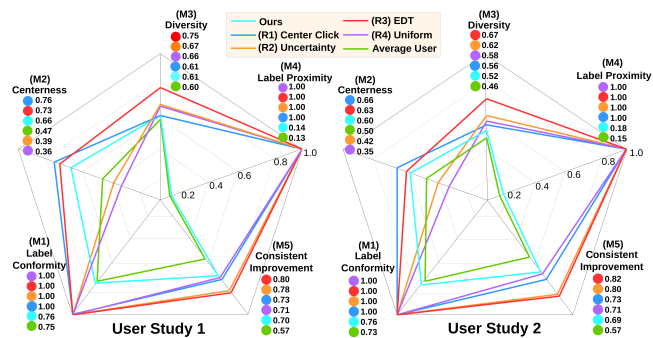}
    \caption{Metric values \textbf{(M1)-(M5)} of all robot users on both user studies.}
    \label{fig:spider-plot}
\end{figure}

\begin{figure}[h]
    \centering
    \begin{subfigure}[b]{0.47\textwidth}
        \includegraphics[width=\textwidth]{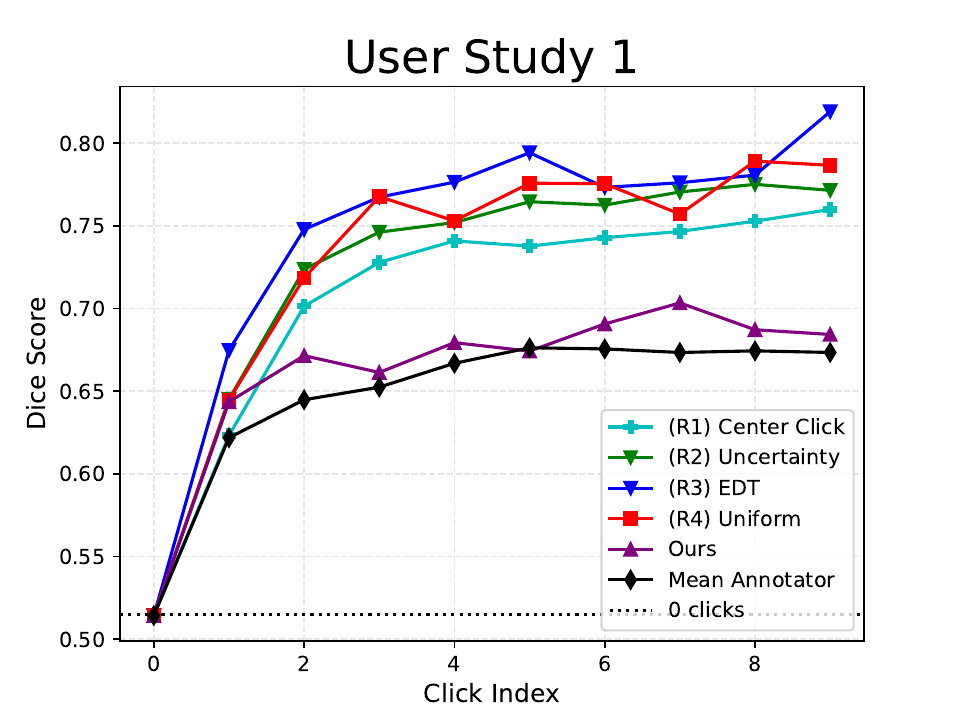}
        \label{fig:dice_us1}
    \end{subfigure}
    \begin{subfigure}[b]{0.47\textwidth}
        \includegraphics[width=\textwidth]{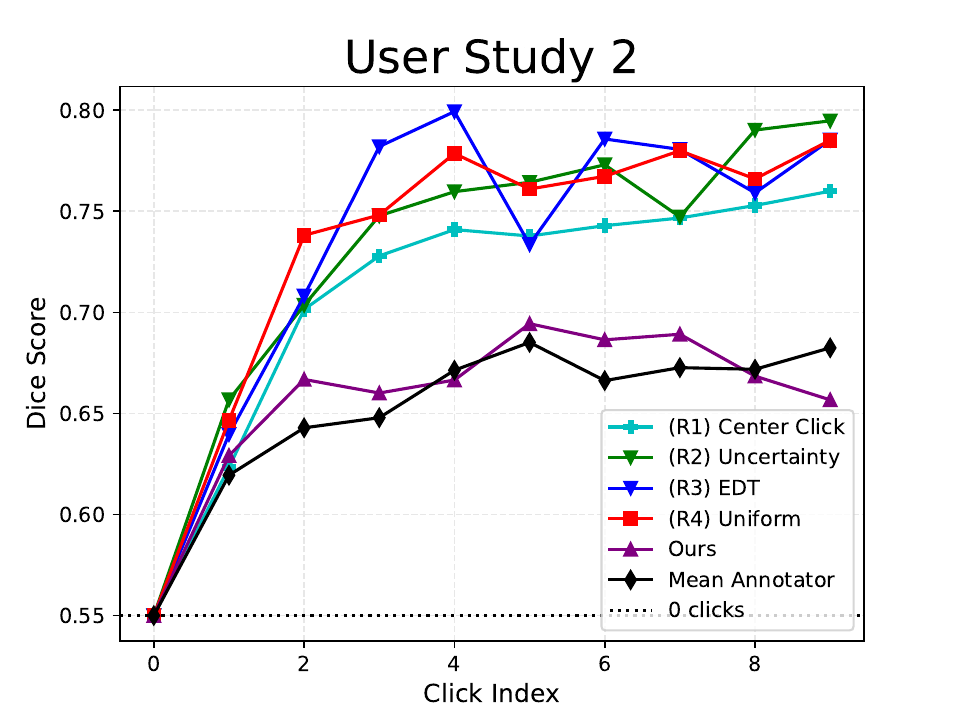}
        \label{fig:dice_us2}
    \end{subfigure}
    \vspace*{-0.8cm}
    \caption{Mean Dice curves of all robot users for both user studies.}
    \label{fig:dice_curves}
\end{figure}

\textbf{Results: Previous Work.} The results, plotted in Fig. \ref{fig:spider-plot} and Table \ref{tab:results} reveal a large discrepancy between existing robot users and the average annotator in all metrics. This contrast is especially notable in \textbf{(M1)} and \textbf{(M4)} since robot users always produce label-conforming clicks, while real annotators click outside the label in $25\%$ of their interactions. Building on this insight, our robot user introduces label non-conformity in $25\%$ of its simulated clicks by spatially perturbing clicks and systematically sampling from high-uptake regions outside the label. This non-conformity achieves the optimal user shift and Dice difference in both user studies. Our robot user reduces the Dice difference from $8.7\%$ to $3.6\%$ and from $7.0\%$ to $3.7\%$ on the first and second user study respectively, which confirms that the Dice score reported when evaluating with our robot user is much more realistic. The Dice curves are visualized in Fig. \ref{fig:dice_curves}.

\vspace*{-0.5cm}

\vspace*{-0.65cm}

\begin{figure}[h]
    \centering
    \includegraphics[width=0.65\textwidth]{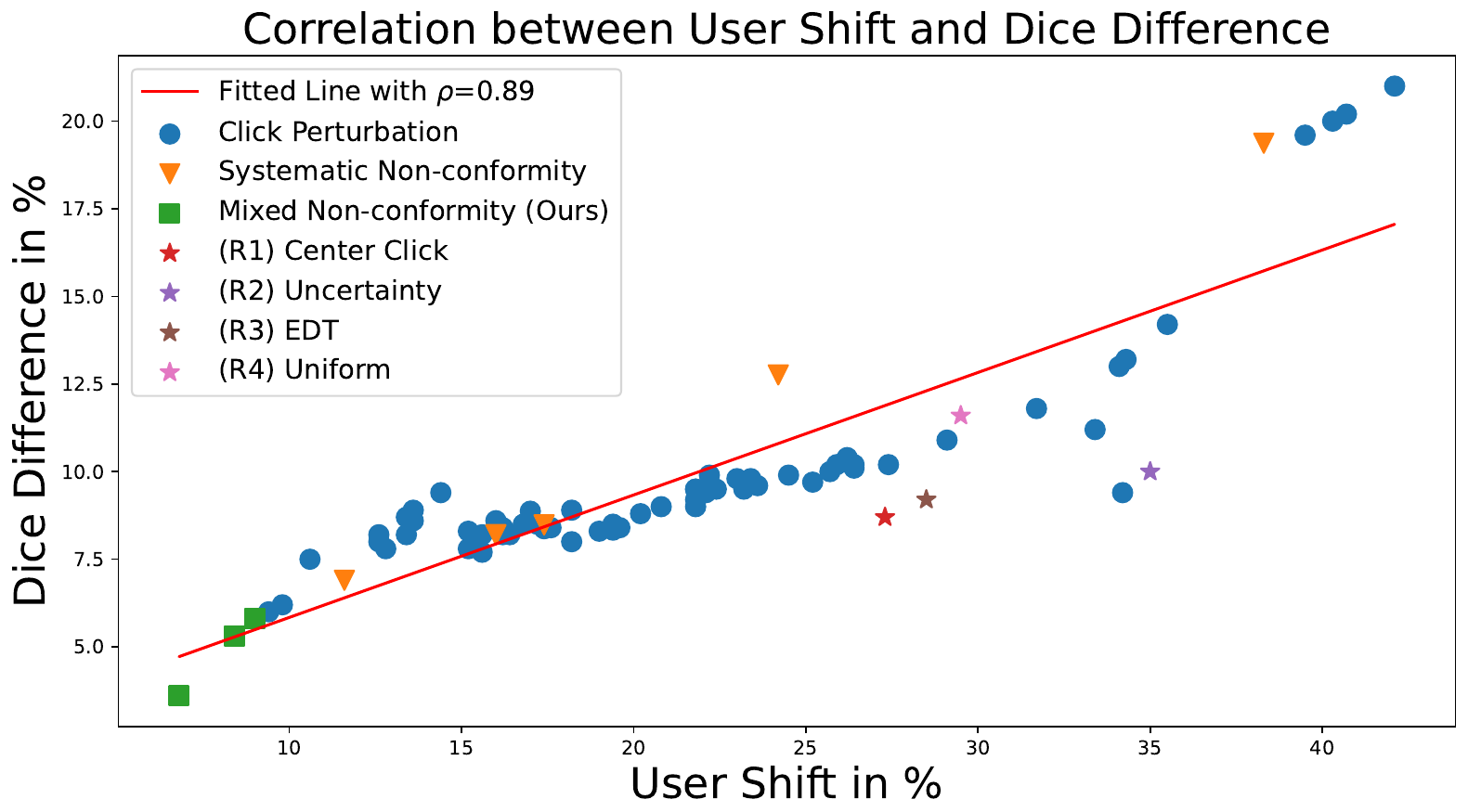}
\vspace*{-0.2cm}
    \caption{Correlation between \textbf{(M6)} and \textbf{(M7)} on the first user study results.}
    \label{fig:correlation}
\end{figure}

\textbf{User Shift vs. Dice Difference.} As the user shift only quantifies the behavioral shift, we examine its correlation with the Dice difference for all our robot user configurations in the first user study. Fig. \ref{fig:correlation} reveals a Pearson correlation of $\rho=0.89$ between the user shift and the Dice difference. Importantly, omitting any of our metrics \textbf{(M1)-(M5)} from \textbf{(M6)} decreases the correlation to $\rho < 0.8$. This confirms that our proposed metrics not only quantify the annotation style but also quantify how this style influences the segmentation performance. 

\vspace*{-0.5cm}

\vspace*{-1cm}
\section{Conclusion}

Our user studies reveal the challenges in evaluating interactive models through simulated interactions. Despite its simplicity, our robot user exposes fundamental flaws in traditional robot users that heavily rely on ground-truth labels. This is particularly problematic in domains where experts disagree on the ground truth in 25\% of their interactions, as observed in our user studies for whole-body PET lesion annotation. Traditional robot users exhibit significant user shift and Dice difference compared to real annotators, resulting in overly optimistic Dice scores and unrealistic annotation behavior. By incorporating click perturbations and systematic label non-conformity, we substantially reduce the user shift and Dice difference compared to previous robot users. This facilitates a more realistic evaluation of interactive model performance without the need for extensive user studies involving the entire test split. 

\textbf{Acknowledgements.}
The user studies were done in collaboration with the Annotation Lab Essen (\url{https://annotationlab.ikim.nrw/}). The present contribution is supported by the Helmholtz Association under the joint research school “HIDSS4Health – Helmholtz Information and Data Science School for Health. This work was performed on the HoreKa supercomputer funded by the Ministry of Science, Research and the Arts Baden-Württemberg and by the Federal Ministry of Education and Research.


%
%
%
%

\end{document}